\documentstyle[prl,aps]{revtex}
\newcommand{\be}{\begin{equation}}
\newcommand{\ee}{\end{equation}}
\newcommand{\bea}{\begin{eqnarray}}
\newcommand{\eea}{\end{eqnarray}}
\newcommand{\rf}[1]{(\ref{#1})}
\newcommand{\tr}{{\rm \, Tr }\, }
\begin{document}
\draft
\title{Quantum Entropy Bound by Information in Black Hole Spacetime}
\author{A.  Hosoya${~}^{\ast}$ and A.  Carlini${~}^{\dagger}$}
\address{$\ast$ Department of Physics, Tokyo Institute of Technology,
Oh-Okayama, Meguro-ku,Tokyo 152, Japan}
\address{$\dagger$ ERATO, Japan Science and Technology Agency, Imai Project,
Bunkyo-ku, Tokyo 113, Japan}
\twocolumn[\hsize\textwidth\columnwidth\hsize\csname
@twocolumnfalse\endcsname \maketitle

\begin{abstract}

    We show that the increase of the generalized entropy by a quantum
    process outside the horizon of a black hole is more than the Holevo
    bound of the classical information lost into the black hole and
    which could be obtained by further observations outside the horizon.  In
    the optimal case, the prepared information can be completely
    retrieved.
\end{abstract}
       \vskip 3ex ]

\narrowtext

\section{Introduction}

Bekenstein \cite{BE1}, on the basis of information theoretical arguments
in a gedanken experiment, proposed the generalized second law in the
black hole spacetime prior to the discovery of the Hawking radiation
\cite{HW} and thus opened up black hole thermodynamics \cite{WA}.
It has been shown that there is an almost complete parallelism between
black hole physics and thermodynamics from the zero-th to the third
law.  However, there remains a long standing problem: the apparent
loss of information about the initial state by evaporation of the
black hole \cite{PAGE}.  From our point of view, it is crucial to
clarify the meaning of ``information" to resolve this paradox.
Recently, the information theoretical aspects in black hole physics
have been reemphasized \cite{BE2} in the light of the entropy bound
conjecture.

In the black hole thermodynamics the total entropy is the sum of the
black hole entropy $S_{BH}=A/4$ (where $A$ is the area of the black
hole horizon, and $S_{BH}=4\pi M^2$ for a spherical black hole of
mass $M$) and of the ordinary matter entropy $S_{M}$, i.e.
$S_{T}=S_{BH}+ S_{M}$.  The generalized second law is motivated by
the paradox of Wheeler's demon: although the entropy $S_{M}$ of the
matter outside the black hole decreases by disposing it to the black
hole, the total entropy $\Delta S_{T}$ increases.  There is plenty of
evidence to support it.  For example, a gedanken experiment suggested
by Unruh and Wald \cite{BE3} takes into account the Unruh effect
\cite{UW1,UW2}, while Frolov and Page \cite{FP} gave a general
argument based on the EPR-like entanglement of the particle states
inside and outside the event horizon.  In a previous work \cite{HCS}
the present authors showed that, in a {\it quantum} version of the
Geroch-Bekenstein gedanken experiment, for the outside region of a
black hole the total entropy increases, while the matter entropy
decreases when a detector is dropped into the black hole.  The
decrease of the matter entropy is more than compensated by the
increase of the black hole entropy via the increase of the black hole
mass which is ultimately attributed to the work done by the
measurement.
In the present work we will show further
that the increase of the generalized entropy is greater than or equal to the
Holevo bound \cite{HOLE,PERE}, which in turn is the upper bound to the
classical information which can be obtained by quantum measurements.
  Entanglement plays an essential role in our argument and
is a key concept of quantum information theory \cite{VE}.

\section{Quantum entropy bound}

The quantum state of the matter in the black hole spacetime is
described by the Hartle-Hawking state,
\be |\psi>_{HH}\equiv
\sum_{n}\sqrt{c_{n}}|n>_{B}|n>_{A},
\label{uno}
\ee
where $c_{n}\equiv \exp[-\omega n/T_{BH}]/Z$ is the Boltzmann factor,
$Z\equiv \sum_{n}\exp[-\omega n/T_{BH}]$ and $T_{BH}\equiv (8 \pi
M)^{-1}$ is the Hawking temperature.  The state (1) is an entangled
state \cite{FP} of the particles inside ($|n>_{B}$) and outside
($|n>_{A}$) of the black hole just like the EPR pair (for a review
see, e.g., \cite{VE}).  The state inside the black hole is not
accessible from the outside so that we trace over the B-state to
obtain a mixed state for the observer outside, i.e.
$\rho_{A}\equiv\tr_{B}(|\psi>_{HH}<\psi|)= \sum_{n}c_{n}|n>_{A}<n|$,
which is nothing but the canonical thermal density operator
\cite{WA2}.  Now imagine a detector of negligible mass in the pure state
$|\Phi_0>$, initially located far away from the black hole horizon,
which is slowly lowered by a string up to a point near the horizon,
and then a quantum experiment outside of the black hole is performed.
The reduced A state will change in general as \be \rho_A \rightarrow
\rho^{\prime}_A\equiv \sum_{\alpha}A_{\alpha}\rho_A
A^{\dagger}_{\alpha}=\sum_{\alpha}p_{\alpha}\rho'_{\alpha},
\label{due}
\ee
with $\sum_{\alpha}A^{\dagger}_{\alpha}A_{\alpha}=1$.  The
transition is represented by a trace preserving positive operator
valued measurement (POVM), where $p_{\alpha}\equiv \tr(A_{\alpha}\rho_A
A^{\dagger}_{\alpha})$ is the probability to get the measurement
result $\alpha$, and $\rho'_{\alpha}\equiv (A_{\alpha}\rho_A
A^{\dagger}_{\alpha})/p_{\alpha}$ is the new normalized density operator.  The
POVM process is more physically understood if we explicitly introduce
detector states $|\Phi_{\alpha}>$ tensored to the entangled state
(\ref{uno}).  In more details, when the agent outside the black hole
switches on his experimental apparatus, the system will undergo a
unitary transformation $U$ for the compound state of A and the
detector as \be |\Psi>\rightarrow |\Psi>',
\label{nove}
\ee
where
\bea
|\Psi>&\equiv
&\sum_{n}\sqrt{c_{n}}|n>_{B}|n>_{A}|\Phi_{0}(x_{0})>
\nonumber \\
|\Psi>'&\equiv
&\sum_{n}\sqrt{c_{n}}|n>_{B}U(|n>_{A}|\Phi_{0}(x_{0})>)
\nonumber \\
&=&\sum_{\alpha, n}\sqrt{c_{n}}|n>_{B}\sum_{m}U^{\alpha}_{nm}|m>_{A}
|\Phi_{\alpha}(x_{0})>,
\label{dieci}
\eea
and where $x_{0}$ is the spacetime point of the detector, which is
initially located outside the horizon.  We assume that by the
measurement the state decoheres (on a proper timescale which ensures
that the process is quasi-static, and which is smaller than the
dynamical timescale of the process itself) to a diagonal form with
respect to the detector states $|\Phi_{\alpha}(x_{0})>$.  The
resultant mixed state $\rho'$ is then
\bea
\rho'&=&\sum_{\alpha}\left (\sum_{n}\sqrt{c_{n}}|n>_{B}\sum_{m}
U^{\alpha}_{nm}|m>_ {A}\right )
\nonumber \\
&\cdot &\left
(\sum_{n'}\sqrt{c_{n'}}_{B}<n'|\sum_{m'}{U^{*\alpha}_{n'm'}}_A<m'|\right
) \nonumber \\
&\otimes &|\Phi_{\alpha}(x_{0})><\Phi_{\alpha}(x_{0})|.
\label{undici}
\eea
However, since the state inside the black hole is not accessible for
the outside observer $A$, we trace over the state of  $B$ to obtain a
reduced density operator for $A$ and the detector as
\bea
\rho_{A\Phi}^{\prime}&\equiv &
\sum_{\alpha}p_{\alpha}\rho^{\prime}_{\alpha}|\Phi_{\alpha}(x_0)>
<\Phi_{\alpha}(x_0)| \nonumber \\
  &=&\sum_{\alpha}A_{\alpha}\rho_A
A_{\alpha}^{\dagger}|\Phi_{\alpha}(x_0)><\Phi_{\alpha}(x_0)|,
\label{dodicibis}
\eea
where $A_{\alpha}\equiv <\Phi_{\alpha}(x_0)|U|\Phi_{0}(x_0)>$.  If
the outside agent does not `read' the detector, the detector states in
eq.  (\ref{dodicibis}) must be traced out and then eq.  (\ref{due}) is
reproduced.  What we have seen above is an explicit construction of a
unitary representation of the POVM where we identify the extended
Hilbert space as that including the detector states \cite{VE}.

Now, the experiment is a local and isothermal process due to the Unruh
effect of the accelerated system with the temperature
$\bar{T}(r)\equiv T_{BH}/\chi(r)$, the blue shifted temperature from
the Hawking temperature $T_{BH}$ of the cavity surrounding the black
hole at infinity.  The first law of black hole physics is \be \Delta
S_{BH}={\Delta M\over T_{BH}}={\Delta W\over T_{BH}},
\label{tre}
\ee
where $\Delta W$ is the work needed for the quantum experiment.  In
the semi-classical gedanken experiment, this corresponds to the work
to push down the box towards the black hole against the buoyancy force
by the Hawking radiation \cite{BE3,UW1,UW2}.

Ordinary thermodynamics tells us that the work $\Delta W$ needed in
the isothermal process is more than or equal to the variation of the free
energy:
\be
\Delta W\geq \Delta F
\label{quattro}
\ee
(with the equality in (\ref{quattro}) holding for a quasi-static
process), where
\bea \Delta F&\equiv
&\sum_{\alpha}p_{\alpha}[E_{\alpha}-\bar{T}(S_{\alpha}'-\log
p_{\alpha})]\chi -(E_{0}- \bar{T}S_{M})\chi \nonumber \\
&=&\left
[S_{M}-\left (\sum_{\alpha}p_{\alpha}S_{\alpha}'-\sum_{\alpha}p_{\alpha}
\log p_{\alpha}\right )\right ]T_{BH},
\label{cinque}
\eea
and we have used the conservation of the internal energy $E_{0}
=\sum_{\alpha}p_{\alpha}E_{\alpha}$, which holds in the isothermal
system ($E_0$ and $E_{\alpha}$ are the energies of the Hawking state
before and after the experiment, respectively).  Furthermore, $S_{M}$
and $S_{\alpha}'$ are defined by $S_{M}\equiv S(\rho_A)$ (the initial
matter entropy) and $S'_{\alpha}\equiv S(\rho'_{\alpha})$, where
$S(\rho)\equiv -\tr(\rho \log\rho) $ is the von Neumann entropy for a
general state $\rho$.  The last term on the r.h.s.  of eq.
(\ref{cinque}) represents the final entropy of the detector, which
reflects our ignorance about the actual outcome of the measurement.

Combining the first law of black hole physics and the second law of
thermodynamics given above, we then easily obtain
$\Delta S_{BH}=S'_{BH}-S_{BH}\geq
S_{M}-\sum_{\alpha}p_{\alpha}(S_{\alpha}'- \log p_{\alpha})$ or, in
a more illuminating way,
\be
(S'_{BH}+S'_{M})-(S_{BH}+S_{M}) \geq 0
\label{sei}
\ee
where $S'_{M}\equiv S(\rho^{\prime}_{A\Phi})=
\sum_{\alpha}p_{\alpha}S'_{\alpha}-\sum_{\alpha}p_{\alpha}\log
p_{\alpha}$ is the matter entropy after the measurement (including the
contribution from the detector).  In other words, the generalized
second law holds.

Let us now extend the previous argument to the case in which the observer
disposes of the detector in a gedanken experiment a l\'{a}
Geroch-Bekenstein.  Suppose that the observer conditionally drops the
detector into the black hole if the experiment outcome is $\alpha\in
D$, while keeping it outside the black hole if $\alpha\notin D$.
That is, the detector might alter the state inside the black hole if the
measurement outcome $\alpha\in D$.  In general the state
(\ref{undici}) will change further to
\bea
\sigma'&=&\sum_{\alpha}\left
(\sum_{n}\sqrt{c_{n}}V_{\alpha}|n>_{B}\sum_{m}
U^{\alpha}_{nm}|m>_{A}\right )
\nonumber \\
&\cdot &\left (\sum_{n'}\sqrt{c_{n'}}_{B}<n'|V_{\alpha}^{\dagger}\sum_{m'}
{U^{*\alpha}_{n'm'}}_A<m'|\right )
\nonumber \\
&\otimes &|\Phi_{\alpha}(x_{\alpha})><\Phi_{\alpha}(x_{\alpha})|,
\label{dodici}
\eea
where $V_{\alpha}$ is a nontrivial unitary transformation if the
experimental outcome is $\alpha\in D$ and $V_{\alpha}=1$ if
$\alpha\notin D$.  Moreover, $x_{\alpha}$ is the spacetime point of
the detector sufficiently after the measurement: $x_{\alpha}$ is
inside the black hole if $\alpha\in D$ and it is outside otherwise.
This corresponds to the ``classical communication from Alice to Bob"
in the standard quantum communication set-up, except that in the present
case it is an inherently one-way communication.

The trace over the B states washes out the $V_{\alpha}$ dependence
altogether and we obtain the reduced density matrix for the compound
state of $A$ and the detector as
\bea
\sigma_{A\Phi}^{\prime}&\equiv & \left (\sum_{\alpha\in
D}p_{\alpha}\rho^{\prime}_{\alpha}\right )\rho_D
\nonumber \\
&+&\sum_{\alpha\not \in D}p_{\alpha}\rho'_{\alpha}
|\Phi_{\alpha}(x_{\alpha})><\Phi_{\alpha}(x_{\alpha})|,
\label{dodiciter}
\eea
where we have introduced the reduced density operator for the detector as
$\rho_D\equiv [\sum_{\alpha\in D}p_{\alpha}
|\Phi_{\alpha}(x_{\alpha})><\Phi_{\alpha}(x_{\alpha})|]/p_D$, with
$p_D\equiv \sum_{\alpha\in D}p_{\alpha}$ the total probability that
the detector is dropped into the black hole.
For $\alpha\in D$ the detector Hilbert space is tensored with the
Hilbert space of the outside observer because the detector and the
outside observer get causally disconnected and therefore decoupled.
It is then straightforward to compute the matter entropy
(now $S'_M\equiv S(\sigma_{A\Phi}^{\prime}$)) as
\bea
S'_{M}&\equiv &
-p_D\sum_{\alpha\in D}\hat p_{\alpha}\log \hat p_{\alpha}
+S\left (p_D\sum_{\alpha\in D}\hat p_{\alpha}\rho^{\prime}_{\alpha}
\right )
\nonumber \\
&+&\sum_{\alpha\not \in D}p_{\alpha}S'_{\alpha}
-\sum_{\alpha\not\in D}p_{\alpha}\log p_{\alpha},
\label{matter2}
\eea
where $\hat p_{\alpha}\equiv p_{\alpha}/p_D$ is the normalized
probability for $\alpha\in D$.

  The change of free energy is still given by eq.  \rf{cinque}, and
an almost identical argument as before leads to
\bea
(&S'_{BH}&+S'_M) -(S_{BH}+S_{M})
\nonumber \\
&\geq& S'_M -\sum_{\alpha}p_{\alpha}S_{\alpha}'
+\sum_{\alpha} p_{\alpha} \log p_{\alpha}.
\label{quindici}
\eea
Finally, substituting eq.  \rf{matter2} into eq.  \rf{quindici} we
obtain

\bea
\Delta S_T&\equiv &(S'_{BH}+S'_M) -(S_{BH}+S_{M})
\nonumber \\
&\geq &p_D\left [S\left (\sum_{\alpha\in D}\hat p_{\alpha}
\rho'_{\alpha}\right ) -\sum_{\alpha\in D}\hat p_{\alpha}S_{\alpha}'
\right ].
\label{quindicibis}
\eea

Now, when the detector is not dropped into the black hole, eq.
\rf{quindicibis} reduces to eq.  \rf{sei}, i.e the generalized second
law holds.  On the other hand, in the dropping case we note that the
quantity inside the brackets on the right hand side of eq.
(\ref{quindicibis}) is the same appearing in the famous Holevo bound
\cite{HOLE,PERE}:
\be
H'_D\equiv  S\left (\sum_{\alpha\in D}
\hat p_{\alpha}\rho_{\alpha}'\right ) -\sum_{\alpha\in D}\hat
p_{\alpha}S(\rho_{\alpha}')\geq I'_D,
\label{sette}
\ee
where $I'_D$ is the mutual information of the components $\alpha
\in D$ which would be obtained if one performed a further observation.
More precisely, with $\{E_{j}\}$ being the orthogonal projection
summing to unity which corresponds to the further observation at
infinity and should be distinguished from the previous POVM, one has
\be
I'_D(E)=-\sum_{j,\alpha\in D}\hat p_{\alpha} p(j|\alpha)\log {p(j)\over
p(j|\alpha)},
\label{otto}
\ee
where $p(j|\alpha)\equiv \tr(E_{j}\rho'_{\alpha})$ is the
conditional probability to obtain the outcome $j$ when the state
$\rho'_{\alpha}$ is prepared and $p(j)\equiv
\sum_{\alpha\in D}\hat p(\alpha)p(j|\alpha)$ is the average probability to
obtain $j$.  Eq.  (\ref{otto}) can be interpreted as the mutual
information between the state prepared by an agent near the black hole
and that of another agent at infinity, i.e.  the uncertainty of the
first measurement minus its uncertainty after the second measurement.
The equality can be achieved for some projection $\{E_{j}\}$ if and
only if the components of the $\rho'_{\alpha}$s are mutually commuting.
In this case the $\rho'_{\alpha}$s can be simultaneously diagonalized
so that we can choose, for example, that
$A^{\dagger}_{\alpha}A_{\alpha}=E_{j}$ as the best that the second
agent can do.  In this optimal case we obtain
$I'_D(E)=-\sum_{\alpha\in D}\hat p_{\alpha}\log \hat p_{\alpha}$, which is
nothing but the Shannon information entropy stored by the first
measurement.
To summarize, eq.  \rf{quindicibis} tells us that, this
potentially acquirable classical information is bounded from above
by the change of the generalized entropy, i.e.
\be
\Delta S_T\geq p_DI'_D
\label{mutual}
\ee
In the ordinary thermodynamics of a closed system $\Delta W=0$, so
that we have $S^{\prime}_{M}-S_M\geq p_DI^{\prime}_D$: the acquirable
information is not more than the change of entropy.

It is also illuminating to consider an ideal case in which the first agent
performs a series of successive quasi-static measurements.  In the
quasi-static isothermal process, the work which is needed under the
influence of an inhomogeneous Hamiltonian $H$ in an experiment a l\'{a}
Stern-Gerlach equals the change of free energy, i.e. $ \Delta W=
\int \tr[\partial_{\bf r} H({\bf r})e^{-\beta H({\bf r})}]\cdot
d{\bf r}/Z=-\beta^{-1}\int \partial_{\bf r} \log Z\cdot d{\bf
r}=\Delta F $, where $Z\equiv \tr[e^{-\beta H({\bf r})}]$ and
$F\equiv -\beta^{-1}\log Z$.  Therefore, the equality is saturated in
eq.  (\ref{mutual}):
\be
\Delta S_T=p_D H'_D=p_D\left [S\left (\sum_{\alpha\in D}\hat
p_{\alpha}\rho'_{\alpha}\right )- \sum_{\alpha\in D}\hat
p_{\alpha}S'_{\alpha}\right ].
\label{diciassette}
\ee
Noting that $H'_D\equiv \sum_{\alpha\in
D}\hat p_{\alpha}S(\rho'_{\alpha}||\rho')$ and the known fact that in
general the relative entropy $S(*||**)$ does not increase by further
measurement \cite{VE}, we see that the amount of increase of the total
entropy becomes less and less at each step of measurement and
eventually does not change at all.  This is reminiscent of Prigogine's
theorem on minimum entropy production \cite{PRIGO}, according to which
the entropy production rate should not increase in a steady state
linear thermodynamical process approaching equilibrium.

Consider a further ideal situation: a quasi-static orthogonal
measurement by the first agent near the black hole followed by the same
orthogonal measurement by the second agent at infinity, so that in eq.
(\ref{mutual}) the equality is doubly saturated, i.e.  $\Delta
S_T=p_DI'_D=-p_D\sum_{\alpha\in D}\hat p_{\alpha}\log \hat p_{\alpha}$,
and a black hole of sufficiently large mass $M$ so that the time scale
of evaporation is slow enough compared with that of the quantum
measurement.  We can then think of the situation where the state
$\sigma'$ is distorted from the thermal state
$\rho_0\equiv|\psi>_{HH}<\psi|$ by the quantum measurement, i.e.
$\rho_0\rightarrow\sigma'$,
and it relaxes back to the initial thermal state $\rho_0$,
assuming that the whole system is surrounded by a cavity with
temperature $T_{BH}$.  When the relaxation $\sigma'\rightarrow\rho_0$
eventually occurs, the energy $\Delta W$ is emitted to infinity in a
form of radiation, and the information $I'_D$ initially stored in the
state $\sigma'$ is encoded in the radiation itself.  Thus, the information
can be completely retrieved by this relaxation process in the ideal
case.  Of course, it is possible to drop matter into a black hole
without distorting the compound state of A and B.  However, in this
case the observer cannot get any information so that he has no
information to loose.  The thermal state remains the thermal state so
that the radiation from the black hole does not carry any information.
\section{Summary and Discussion}

We have shown that the increase of the generalized entropy by
a quantum process outside the horizon of a black hole is more than the
Holevo bound of classical mutual information lost into the black hole.
What we have used as physics are the energy conservation for an
isothermal process in the black hole spacetime and the second law of
ordinary thermodynamics.  The difference between the ordinary POVMs
and those in the black hole spacetime is that the {\it work} needed
for the experiment makes the black hole more massive.  One might
consider ours as a special and hypothetical gedanken experiment.
After a little thought, however, one may realize that this represents
a fact of real life.  After all black holes exist somewhere in the
universe and any physical process can be considered as a POVM outside
the black holes.  The present argument is universal not only in the
sense that POVMs represent the most general physical process
including, for example, gas collision before the infall, but also in
the sense that the quantum state is entangled for all kinds of
particles because gravity is universally coupled to any matter.  Of
course our discussion does not completely solve the information loss
paradox, because our treatment of the black hole is semi-classical.
One will need a full theory of quantum gravity to really understand
the process of information loss and retrieval after a complete
evaporation of the black hole, the final stage of which is expected
to be trans-Planckian.

In conclusion, our suggestion is that the information loss paradox is not
merely an issue of evolution from pure to mixed states, but rather it
should be fully addressed within the context of quantum measurement
and information theory.

\bigskip

\noindent {\Large \bf Acknowledgements}

\bigskip
A.H.'s research was partially supported by the Ministry of Education,
Science, Sports and Culture of Japan, under grant n. 09640341.

\end{document}